\documentclass[prd,amsmath,amssymb,floatfix,nofootinbib,preprintnumbers,twocolumn]{revtex4}
\usepackage[latin1]{inputenc}
\usepackage{amsmath}
\usepackage{amsfonts}
\usepackage{hyperref}
\hypersetup{colorlinks=True,citecolor=blue}
\usepackage{amssymb}
\usepackage{graphicx,multirow}
\usepackage{appendix}
\usepackage{dcolumn}
\usepackage{caption}
\usepackage{subfigure}
\usepackage{float}
\usepackage{bm}
\usepackage{latexsym}
\usepackage{epsfig}
\usepackage{graphicx,color}
\newcommand{\cg}[6]{\mathcal{C}^{#1 #2}_{#3 #4 #5 #6}}
\newcommand{\ylm}[2]{Y_{#1 #2}}        
\newcommand{\ylms}[3]{\; {}_{#3}Y_{#1 #2}}   
\newcommand{\ncap}{\mathbf{\hat{n}}}
\newcommand{\al}{\text{\large{$\alpha$}}}  
\newcommand{\be}{\begin{equation}}
\newcommand{\ee}{\end{equation}}
\newcommand{\ba}{\begin{eqnarray}}
\newcommand{\ea}{\end{eqnarray}}
\newcommand{\nn}{\nonumber \\}

\begin{document}
\title{Weak lensing in non-statistically isotropic universes}
\author{Moumita Aich$^{1}$\footnote{aich@ukzn.ac.za}, Aditya Rotti$^{2,3}$\footnote{adityarotti@gmail.com} and Tarun Souradeep$^{3}$\footnote{tarun@iucaa.ernet.in}}
\address{ $^1$Astrophysics and Cosmology Research Unit, School of Mathematics,
Statistics \& Computer Science, University of KwaZulu-Natal, Durban - 4001, South Africa \\ 
$^2$ Florida State University, Tallahassee, FL, U.S.A\\
$^3$IUCAA, Post Bag 4, Ganeshkhind, Pune-411007, India\\}
\begin{abstract}
The Bipolar Spherical Harmonics (BipoSH) form a natural basis to study the CMB two point correlation function in a non-statistically isotropic (non-SI) universe.  The coefficients of expansion in this basis are a generalization of the well known CMB angular power spectrum and contain complete information of the statistical properties of a non-SI but Gaussian random CMB sky. We use these coefficients to describe the weak lensing of CMB photons in a non-SI universe. Finally we show that the results reduce to the standard weak lensing results in the isotropic limit.
\end{abstract}
\maketitle	
\section{Introduction}
We assume the CMB temperature anisotropies to be Gaussian, which is in good agreement with current CMB observations. Consequently the two point correlation function contains complete information about the underlying CMB temperature field. Generically, the two point correlation function can be expressed in terms of the spherical harmonic coefficients of CMB temperature maps as follows,
\ba \label{corrfn}
	C(\mathbf{\ncap_1},\mathbf{\ncap_2})&=& \langle\Delta
T(\mathbf{\ncap_1})\Delta
T(\mathbf{\ncap_2})\rangle,\nn
	&=& \sum_{lml'm'} \langle a_{lm}a^*_{l'm'}\rangle Y_{lm}(\mathbf{\ncap_1})
Y_{l'm'}^{*}(\mathbf{\ncap_2})	\,,
\ea
where $\langle a_{lm}a^*_{l'm'}  \rangle$ is commonly referred to as the covariance matrix. 

Now, if one assumes the universe to be statistically isotropic (SI), it can be argued that the two point correlation function of the CMB temperature field cannot have any explicit dependence on the directions $\ncap_1$ and $\ncap_2$ and hence can only depend on the angular separation between the two directions,
\ba \label{cl1}
C(\mathbf{\ncap_1},\mathbf{\ncap_2})&=&C(\mathbf{\ncap_1} \cdot \mathbf{\ncap_2}) = C(\theta) \,,
\ea
where $\theta = cos^{-1}(\hat{n}_1 \cdot \hat{n}_2)$ is the angular distance between the two directions.
This correlation function can be expanded in the Legendre polynomial $P_l$ basis as follows,
\ba\label{cl2}
	C(\mathbf{\ncap_1},\mathbf{\ncap_2})&=& \sum_l \frac{2l+1}{4\pi} C_{l} P_{l}(\mathbf{\ncap_1} \cdot
\mathbf{\ncap_2}) \,, 
\ea
where CMB angular power spectrum $C_l$ is the coefficient of expansion in this basis. It can be shown that $C_l$ is related to the covariance matrix through the following expression,
\ba
	C_l&=& \langle a_{lm}a^*_{l'm'}\rangle \delta_{l l'}\delta_{mm'} \,,
\label{unlensed-cl}
\ea
implying that in a SI universe the complete statistical information on the CMB field is encoded in the diagonal elements of the covariance matrix.
\section{Describing a non-statistically isotropic universe}
If we make no assumption about a SI universe, then the two point correlation function explicitly depends on the directions $\ncap_1$ and $\ncap_2$. This property reflects itself in harmonic space, as non-vanishing off-diagonal elements in the covariance matrix. This implies that the angular power spectrum by itself cannot fully characterizes the statistical properties of a non-SI CMB sky. In fact, searches for non-vanishing off-diagonal elements in the covariance matrix forms the basis of all tests searching for violations of isotropy.

In a non-SI universe the two point correlation is best expressed in terms of a basis which has this explicit bi-directional dependence. The natural choice is then that of the Bipolar Spherical Harmonic (BipoSH) functions \cite{AH-TS, AH-TS1},  which form a complete orthonormal basis for functions defined on $S^2 \times S^2$. The two point correlation function for any non-SI field can be expanded in the BipoSH basis as follows,
\be
C(\ncap_1,\ncap_2) = \sum_{LM l_1 l_2} A^{LM}_{l_1 l_2} \{ Y_{l_1}(\ncap_1) \otimes Y_{l_2}(\ncap_2) \}_{LM} \,.
\label{gen-2pt}
\ee
where $A^{LM}_{l_1 l_2}$ are the coefficients of expansion in the BipoSH basis. The BipoSH basis functions themselves, can be expressed in terms of the standard spherical harmonic functions $Y_{lm}$ as,
\be
\{ Y_{l_1}(\ncap_1) \otimes Y_{l_2}(\ncap_2) \}_{LM} = \sum_{m_1 m_2} \cg L M {l_1} {m_1} {l_2} {m_2} \ylm{l_1}{m_1}(\ncap_1) \ylm{l_2}{m_2}(\ncap_2) \,,
\label{biposh-basis-def}
\ee
where $\cg L M {l_1} {m_1} {l_2} {m_2}$ are the Clebsch-Gordon coefficients. We refer to $L$ as BipoSH multipoles and $l_1, l_2$ as spherical harmonic (SH) multipoles. The Clebsch-Gordon coefficients are defined only when their indices satisfy the following constraint equations,
\begin{subequations}
\ba
 |l_1-l_2| \leq L \leq l_1+l_2 \,, \\
 m_1 + m_2 = M \,, \\
 -l_1 \leq m_1 \leq l_1 \,, \\
 -l_2 \leq m_2 \leq l_2 \,.
 \ea
 \end{subequations}
Using the orthonormality property of the BipoSH basis functions, it can be shown that the BipoSH spectra $A^{LM}_{l_1 l_2}$ are related to the the covariance matrix as follows,
\ba
 A^{LM}_{l_1 l_2} = \sum_{m_1 m_2} \langle a_{l_1 m_1}a_{l_2 m_2} \rangle \cg L
M {l_1} {m_1} {l_2} {m_2} \,.
\ea 
As evident from the above equation, all the information in the covariance matrix is captured in the BipoSH spectra. Specifically it can be shown that  the BipoSH spectrum $A^{00}_{ll}$ is related to the CMB angular power spectrum as follows,
\be \label{iBipoSH}
	A^{00}_{ll}=(-1)^l \Pi_l C_l, 	
\ee
where $\Pi_l=\sqrt{2l+1}$. Therefore, the BipoSH spectra are a generalization to the commonly studied CMB angular power spectrum. To summarize, while the  BipoSH spectrum $A^{00}_{l l}$ characterize the isotropic component of the field, the rest of the BipoSH spectra $A^{LM}_{l_1 l_2} (~L \neq 0)$, describe the non-SI component of the field. The primary scheme to detect a violation of isotropy in the data has been to search for these non-vanishing BipoSH spectra with $L>0$ \cite{CB-RH-GH,planck2013_13_I&S,PhysRevD.91.043501,planck2015}.

Before we proceed, we discuss a few more interesting properties of the BipoSH spectra. For any fixed $l_1,l_2$ the BipoSH spectra $A^{LM}_{\cdot \cdot}$ have the same mathematical properties as the spherical harmonic coefficients of a map. Motivated by this property it is possible to define the reduced BipoSH map formed by summing over the SH multipoles of the BipoSH spectra \cite{AH-TS2},
\be \label{bipolarmap}
\mathcal{R}^{LM}=\sum_{l_1 l_2}  A^{LM}_{l_1 l_2}.
\ee
It is also possible to define the Bipolar power spectrum (BiPS), a coordinate independent quantity which is a biased estimate of SI \cite{AH-TS} and is defined as 
\be \label{bipolarspectrum}
\kappa_L=\sum_{l_1,l_2,M} \vert A^{LM}_{l_1 l_2} \vert^2 \,.
\ee
We will discuss how these completely generically defined quantities are related to some cosmological observables in a later section of this article. 

\section{Weak lensing of the CMB temperature field}
The CMB photons reach the observer after traversing through the intervening large-scale structures (LSS). This results in the CMB photons deviating from their geodesics in the unperturbed Friedmann-Lemaitre-Robertson-Walker (FLRW) metric. We work with the approximation that the deviations from the geodesics are small, such that the photons sample the same gravitational potentials as they would have along the unperturbed trajectories, termed the Born approximation \cite{Challinor2002}. 

Weak lensing of the CMB photons results in the CMB temperature map getting remapped,
\ba \label{remap}
\tilde{T}(\ncap)&=& T(\ncap+\vec{\Delta}) \,,
\ea
where $\vec{\Delta}$ is the deflection to the direction of photon arrival. On solving the photon geodesic equation under the Born approximation it can be show that $\vec{\Delta}$ is related to the projected lensing potential $\psi$  as follows, 
\ba \label{deltapsi}
\vec{\Delta}(\ncap)=\vec{\nabla} \psi(\ncap) \,.
\ea
The projected lensing potential $\psi$ is expressed as a weighted sum of the gravitational potential encountered by the photon along its trajectory,
\ba
\psi(\ncap)=\int_{\eta_0}^{\eta_s} d\eta \left[\frac{\eta_0-\eta}{\eta_0 \eta}\right] \phi(\eta,\ncap),
\ea
where $\eta$ is the angular diameter distance which is identical to the comoving distance in a flat universe and $\phi(\eta,\ncap)$ is the gravitational potential encountered by the photon coming in from direction $\ncap$ at a comoving distance of $\eta$. 

Analogous to the CMB temperature field the projected lensing potential is a Gaussian random field defined on a sphere and thus in harmonic space 
\ba  
\psi(\ncap)&=&\sum_{kn} \psi_{kn} \ylm{k}{n}(\ncap) \;, \\
C^{\psi \psi}_{k} &=& \langle \psi_{kn}\psi^*_{k'n'} \rangle \delta_{k k'} \delta_{n n'} \;, 
\ea
where $\psi_{kn}$ denotes the spherical harmonic coefficients of the lensing potential and $C^{\psi \psi}_{k}$ the corresponding angular power spectrum in a SI universe. 

In a non-SI universe the statistical properties of $\psi$ are characterized by the following BipoSH spectra,
\be
\Psi^{KN}_{k_1 k_2} = \sum_{n_1 n_2} \langle \psi_{k_1 n_1} \psi_{k_2 n_2} \rangle \; \cg K N {k_1} {n_1} {k_2} {n_2} \, .
\ee
where the BipoSH coefficient $\Psi^{KN}_{k_1 k_2}$ ($\forall~K \ne 0$) encodes the information in the off-diagonal components of the harmonic space covariance matrix of the projected lensing field.

Weak lensing of the CMB photons through the intervening LSS leave measurable imprints on the CMB two point correlation function. Specifically lensing introduces coupling between different multipole moments of the temperature field, which are absent in an unlensed SI CMB sky (See Eq. \ref{unlensed-cl}). The lensing modification to the CMB angular power spectra (the diagonal components of the covariance matrix) have been well studied \cite{US1,WH}. 

In the following sections of this article we discuss weak lensing induced modifications to the BipoSH spectra. 
\section{BipoSH representation of a lensed CMB sky}\label{tlenbips}
Since we measure the lensed CMB sky, while modeling the two point correlation we must account for the weak lensing induced modifications to the statistics of the map. As discussed in the previous section, lensing remaps the temperature field. Given the statistics of the projected lensing potential $\psi$, it is possible to evaluate the lensing modifications to the BipoSH spectra.


Since the deflection $\vec{\Delta}$ is small, we can Taylor expand and express the lensed temperature field $\tilde{T}(\ncap)$ in terms of the unlensed temperature field $T(\ncap)$ and the deflection field $\vec{\Delta}(\hat{n})$. In order to evaluate the lensing modifications to the two point correlation of the CMB temperature field we retain terms which are at most second order in deflections.  Hence to leading order terms in the deflection field $\vec{\Delta}(\hat{n})$, the lensed temperature field $\tilde{T}$ can be expressed in terms of the unlensed temperature field $T$ as follows,
\ba
\tilde{T}(\ncap)&=& T(\ncap+\Delta) , \nn
&\approx& T(\ncap)+\nabla^a \psi(\ncap) \nabla_a T(\ncap) \nn &+& \frac{1}{2} \nabla^a\psi(\ncap)
\nabla^b\psi(\ncap) \nabla_a \nabla_b T(\ncap) +  O(\psi^3)\,.
\ea
In the above equation we have used the fact that the deflection field $\vec{\Delta}(\hat{n})$ is related to the gradient of the projected lensing potential. Consequently, the two point correlation of the lensed temperature field can be expressed as follows,
\begin{widetext}
\ba \label{2ptcorr}
	\langle \tilde{T}(\ncap_1)\tilde{T}
(\ncap_2) \rangle &=&\langle T(\mathbf{\ncap_1}+\mathbf{\Delta_1})T(\mathbf{\ncap_2}+\mathbf{\Delta_2})\rangle	 \nn
	&=& \langle T(\mathbf{\ncap_1})T(\mathbf{\ncap_2})\rangle + \langle\nabla^a \psi(\mathbf{\ncap_1}) 
	\nabla_a T(\mathbf{\ncap_1})T(\mathbf{\ncap_2})\rangle +\langle\nabla^a \psi(\mathbf{\ncap_2}) 
	\nabla_a T(\mathbf{\ncap_2})T(\mathbf{\ncap_1})\rangle \nn
	&+& \langle \nabla^a\psi(\ncap_1) \nabla^b \psi(\ncap_2) \nabla_a T(\ncap_1) \nabla_b T(\ncap_2) \rangle \nn 
	&+& \frac{1}{2}\langle \nabla^a\psi(\ncap_1) \nabla^b \psi (\ncap_1) \nabla_a \nabla_b T(\ncap_1)T(\ncap_2) \rangle 
	+ \frac{1}{2}\langle \nabla^a \psi (\ncap_2) \nabla^b \psi (\ncap_2) \nabla_a \nabla_b T(\ncap_2)
T(\ncap_1) \rangle \;.
\ea
\end{widetext}
Here we follow the Einstein summation convention, i.e., repeated indices of sky coordinates are summed over.

We make a small diversion to discuss some subtle assumptions we have made in the analysis that follows. 
\begin{itemize}
\item We emphasize the point that unlike the CMB temperature field, we do not treat the projected lensing potential as a stochastic field, instead we treat it as a constant field around an observer. This is justified by the fact that we are surrounded by a particular realization of the LSS and we see the CMB through this particular realization. 
\item Since the averaging is done over a single realization, the term linear in the projected lensing potential $\psi$ does not vanish. 
\end{itemize}

The remarks made above are very critical to the analysis presented in this article. Since the dominant contribution to the projected lensing potential comes from very local structures, any anomaly in the local distribution of structures around us would immediately reflect in the CMB observations. We demonstrate in the following sections that the BipoSH coefficients are observables constructed from observed CMB maps which can in principle allow us to test the nature of the local universe.

The operation of taking the gradient transverse to the line-of-sight  ($\nabla^a \equiv \nabla^a_{\ncap}$) and the averaging $\langle...\rangle$ operation commute and we use this property to express the lensed two point correlation function of the temperature field as follows,
\begin{widetext}
\ba
	\langle \tilde{T}(\ncap_1)\tilde{T}(\ncap_2) \rangle &=& \langle T(\ncap_1)T(\ncap_2) \rangle + \nabla^a \psi(\ncap_1) \nabla_a^{\ncap_1}
	\langle T(\ncap_1)T(\ncap_2) \rangle  + \nabla^a \psi(\ncap_2) \nabla_a^{\ncap_2}
	\langle T(\ncap_1)T(\ncap_2) \rangle \nn
	&+& \nabla^a_{\ncap_1} \nabla^b_{\ncap_2} \langle \psi(\ncap_1) \psi(\ncap_2) \rangle  
	\nabla_a^{\ncap1} \nabla_b^{\ncap2} \langle T(\ncap_1) T(\ncap_2) \rangle 
	+\frac{1}{2} \left[ \nabla^a_{\ncap_1} \nabla^b_{\ncap_1} \langle \psi(\ncap_1)\psi(\ncap_1) \rangle 
	\nabla_a^{\ncap1} \nabla_b^{\ncap1} \langle T(\ncap_1) T(\ncap_2) \rangle \right. \nn &+&  \left. \nabla^a_{\ncap_2} \nabla^b_{\ncap_2} \langle \psi(\ncap_2)\psi(\ncap_2) \rangle \nabla_a^{\ncap_2} \nabla_b^{\ncap_2} \langle T(\ncap_2) T(\ncap_1) \rangle \right] . \,\,
\label{2ptlens}
\ea
\end{widetext}
where the gradient now operates on the ensemble averages of the field, which can be expressed in the BipoSH basis. Note that the terms which are linear in $\psi$ are written explicitly without the ensemble averaging. We reiterate that while dealing with terms which are first order in $\psi$ the projected lensing potential is not treated as a stochastic field, instead it is thought of as a constant field. 

For notational convenience, we identity each of the terms in the two point correlations in the above equation as,
\ba
\tilde{A} &=& A + (\al_1 + \al_2) + \beta + (\gamma_1 + \gamma_2) \,, \nn 
	&=& A + \al + \beta + \gamma  \,,
\ea
where $\alpha$ denotes the terms linear in the lensing potential $\psi$, $\beta$ denotes terms which are quadratic in $\psi$ but consist of only first derivative of the temperature field and $\gamma$ denotes terms which are quadratic in $\psi$ but consist of  double derivative of the temperature field. Next, we express Eq.~\ref{2ptlens} in harmonic space, by expressing each two point correlation in the BipoSH basis. The transverse gradients now operate on these basis functions. A detailed demonstration of this procedure can be found in Appendix \ref{app:cov}. 

Finally this relates the BipoSH coefficients $\tilde{A}^{LM}_{l_1 l_2}$  of the lensed temperature field to the unlensed temperature BipoSH coefficients $A^{LM}_{l_1 l_2}$ through the BipoSH coefficients for the lensing deflection field $\Psi^{LM}_{l_1 l_2}$,
\ba
	\tilde{A}^{LM}_{l_1 l_2}&=& A^{LM}_{l_1 l_2}+  \left[ \;_1\al^{LM}_{l_1l_2} + \;_2\al^{LM}_{l_1l_2} \right] + \beta^{LM}_{l_1 l_2} \nn &+&
	  \left[ \;_1\gamma^{LM}_{l_1l_2} + \;_2\gamma^{LM}_{l_1l_2} \right] \,, \nn
	  &=&A^{LM}_{l_1 l_2}+ \;\al^{LM}_{l_1l_2}(\psi_{kn},A^{L'M'}_{l_1 l_2})
	+ \beta^{LM}_{l_1 l_2}(\Psi^{KN}_{k_1 k_2},A^{L'M'}_{l_3 l_4}) \nn &+& \gamma^{LM}_{l_1 l_2}(\Psi^{KN}_{k_1 k_2},A^{L' M'}_{l_3 l_4}),
\label{lens-biposh}
\ea
where $\al^{LM}_{l_1l_2}$ is the lensing correction term which is first order in $\psi$ and ($\beta^{LM}_{l_1 l_2}, \; \gamma^{LM}_{l_1 l_2}$) are the lensing correction terms which are second order in $\psi$. 

In the following sections we evaluate each of these terms explicitly and discuss their consequences.


\subsection{First order correction terms}
We begin by evaluating the lensing correction to the BipoSH coefficient arising due to the term which is linear in the lensing field $\psi(\hat{n})$. The linear order correction is given by the following expression,
\ba
\al^{LM}_{l_1 l_2} &=& \psi_{LM} \frac{C_{l_1}F(l_1,L,l_2)+C_{l_2}F(l_2,L,l_1)}{\sqrt{4\pi}} \frac{\Pi_{l_1}\Pi_{l_2}}{\Pi_L} \cg L 0 {l_1} 0 {l_2} 0 \nn
&+&\sum_{kn} \psi_{kn} \sum_{m_1 m_2} \cg {L}{M} {l_1}{m_1}{l_2}{m_2} \sum_{L' M' l_3 m_3}^{L' \neq 0} \left[A^{L' M'}_{l_3 l_2} \cg {L'}{M'} {l_3}{m_3}{l_2}{m_2} \right. \nn & \times & \left. I^{m_1 n m_3}_{l_1 k l_3}  +   
A^{L' M'}_{l_1 l_3} \cg {L'}{M'} {l_1}{m_1}{l_3}{m_3} I^{m_2 n m_3}_{l_2 k l_3} \right], 
\label{alpha}
\ea
where the first term is assuming the underlying unlensed temperature field to be SI i.e. $L'\neq 0$ while the second term is the non-SI term. The integral $ I^{m_1 n m_3}_{l_1 k l_3}$ in the above equation can be solved to yield the following expression,
\ba
I^{m_1 n m_3}_{l_1 k l_3} &=& \int d\ncap_1 ~\ylm{k}{n}^{:a} \ylm{l_3}{m_3:a} \ylm{l_1}{m_1}^* \nn 
&=& \frac{F(l_3, k, l_1)}{\sqrt{4\pi}} \frac{\Pi_k \Pi_{l_3}}{\Pi_{l_1}} \cg {l_1} 0 k 0 {l_3} 0 \cg {l_1} {m_1} k n {l_3} {m_3} \;,
\label{i-int}
\ea
where $F(l_3, k, l_1) = [l_3(l_3+1)+k(k+1)-l_1(l_1+1)]/2$. 

The first thing to note in Eq.~\ref{alpha} is that this term vanishes on taking an average over the lensing field $\psi_{kn}$. However while making observations of the CMB sky we are surrounded by a single realization of the large scale structure and hence in practice it is this single realization of the LSS that biases our observations.

In Eq. \ref{alpha}, the first term results from the isotropic component of the CMB sky whose statistical properties are fully specified by the angular power spectrum. Examining the first term also yields the fact that the BipoSH coefficients derived from the lensed CMB sky, even in a statistically isotropic universe, will be rendered non-zero due to weak lensing. While the amplitude of the BipoSH spectra is determined by the magnitude of the projected lensing potential, the spectral shape of the signal is completely determined by the CMB angular power spectra.
It can also be seen that there are no corrections to the power spectrum arising from this term, since $\alpha^{00}_{ll}$ identically vanishes.

The second term describes the coupling between various BipoSH spectra due to weak lensing. It is not possible to extract much information from this term without assuming a particular model of isotropy violation. 

\subsubsection*{Estimators for lensing variables.}
As is evident from the above discussion, the BipoSH coefficients encode information about the projected lensing potential. It is possible to derive an estimator for the projected lensing field and the corresponding power spectra in terms of quantities derived from BipoSH coefficients which can be measured from observations of CMB temperature maps. The projected lensing potential is expressed in terms of the bipolar maps defined in Eq.~\ref{bipolarmap} through the following expression,
\ba
\hat{\psi}_{LM}&=&\frac{\sqrt{4\pi} \Pi_L \mathcal{R}^{LM}}{\displaystyle{\sum_{l_1 l_2}}\left[ C_{l_1}F(l_1,L,l_2)+C_{l_2}F(l_2,L,l_1)\right] C^L_{l_1 l_2}} \, ,
\label{lens-estm}
\ea
where $C^L_{l_1 l_2} =\Pi_{l_1}\Pi_{l_2} \cg L 0 {l_1} 0 {l_2} 0 $.

Similarly it can be shown that the power spectrum of the projected lensing potential is related to the BiPS $\kappa_L$ defined in Eq.~\ref{bipolarspectrum}, through the following expression,
\ba
\hat{C}_{L}^{\psi\psi} = \frac{4\pi \kappa_L}{\displaystyle{\sum_{l_1 l_2}}\left[ C_{l_1}F(l_1,L,l_2)+C_{l_2}F(l_2,L,l_1) C^L_{l_1 l_2} \right]^2} \, .
\label{ps-estm}
\ea
Note that $\kappa_L$ is a biased estimate of the lensing power spectrum. Hence studying the two point correlation of the CMB temperature field in the BipoSH basis very naturally leads to estimators for the lensing variables (the lensing field $\psi_{LM}$ and the lensing field power spectrum $C_L^{\psi\psi}$). These constructions are similar to the lensing estimators proposed in \cite{okamoto-hu}, though the ones presented here are not optimal in the sense that they do not minimize the variance on the reconstructed lens harmonics $\psi_{LM}$.

The above discussion is fully valid for lensing reconstruction in a SI universe. This scheme of lensing reconstruction is approximate for a non-SI universe. For this scheme to be valid in a non-SI universe it is necessary that the second term in Eq. \ref{alpha} is significantly smaller than the first. This may be the case since $C_l$ is always larger than the other BipoSH spectra in a non-SI universe. While in a SI universe the BipoSH spectra measured from the lensed CMB sky can be directly used to infer the lensing potential, in a non-SI universe it will be necessary to subtract the BipoSH spectra of the unlensed CMB sky from the measured BipoSH spectra in order to estimate the lensing potential.
\subsection{Second order correction terms}
We now evaluate the other correction terms to the lensed temperature BipoSH coefficients in Eq. \ref{lens-biposh} which are $\beta^{LM}_{l_1 l_2}$ and $\gamma^{LM}_{l_1 l_2}$. Both these terms are second order in the projected lensing potential $\psi$. In the most general scenario, these terms are functions of the BipoSH coefficients of the projected lensing potential and that of the unlensed temperature field. Interestingly, the $\gamma$ term, like the first order correction term $\alpha$, comprises of two symmetrical terms arising from ensemble averaging along all line of sights. However, because of the complexity of the second order contribution of the lensing potential, the integral in the $\gamma$ term is more involved. 
\begin{widetext}
\begin{subequations}
\ba
\label{beta}
\beta^{LM}_{l_1 l_2} &=& \sum_{K N k_1 k_2} \Psi^{KN}_{k_1 k_2} \sum_{n_1 n_2} \cg {K}{N}{k_1}{n_1}{k_2}{n_2} 
\sum_{L' M' l_3 l_4} A^{L' M'}_{l_3 l_4} \sum_{m_3 m_4} \cg {L'}{M'} {l_3}{m_3}{l_4}{m_4} 
\sum_{m_1 m_2}  \cg {L}{M}{l_1}{m_1}{l_2}{m_2} I^{m_1 n_1 m_3}_{l_1 k_1 l_3} I^{m_2 n_2 m_4}_{l_2 k_2 l_4} \,, \\
\gamma^{LM}_{l_1 l_2} &=& _1\gamma^{LM}_{l_1 l_2} + _2\gamma^{LM}_{l_1 l_2} \,, \nn
      &=& \frac{1}{2}\sum_{K N k_1 k_2} \Psi^{KN}_{k_1 k_2} \sum_{n_1 n_2} \cg {K} {N} {k_1} {n_1} {k_2} {n_2}
      \sum_{m_1 m_2} \cg {L} {M} {l_1} {m_1} {l_2} {m_2}  \nn
      &\times& \sum_{L' M' l_3 m_3} \left[ A^{L' M'}_{l_3 l_2} \cg {L'} {M'}{l_3} {m_3} {l_2} {m_2} J^{m_1 n_1 m_3 n_2}_{l_1 k_1 l_3 k_2} + 
      A^{L' M'}_{l_3 l_1} \cg {L'} {M'}{l_3} {m_3} {l_1} {m_1} J^{m_2 n_1 m_3 n_2}_{l_2 k_1 l_3 k_2} \right] \; ,     
\label{gamma}
\ea
\end{subequations}
where the term $J^{m_1 n_1 m_3 n_2}_{l_1 k_1 l_3 k_2}$ is the following integral,
\ba
J^{m_1 n_1 m_3 n_2}_{l_1 k_1 l_3 k_2} &=&
\int  d{\ncap_1} \ylm{l_1}{m_1}^*(\ncap_1) \ylm{k_1}{n_1}^{:a}(\ncap_1) \ylm{k_2}{n_2}^{:b}(\ncap_1) \ylm{l_3}{m_3 \: :ab}(\ncap_1) \, .
\label{j-int}
\ea
The evaluation of this integral is illustrated in Appendix \ref{app:int}. This integral can be explicitly evaluated in terms of Clebsch-Gordon coefficients and is given by the following expression,
\ba
J^{m_1 n_1 m_3 n_2}_{l_1 k_1 l_3 k_2} &=& \frac{a_{k_1} a_{k_2} a_{l_3}}{4\pi} \frac{\Pi_{k_1} \Pi_{k_2} \Pi_{l_3} }{\Pi_{l_1}} 
\sum_{kn} \cg k n {k_1} {n_1} {k_2} {n_2} \cg {l_1} {-m_1} {k} {-n} {l_3} {-m_3} \nn &\times & \left[ a_{l_3} ~ \cg k 0 {k_1} 1 {k_2} {-1} ~ \cg {l_1} 0 k 0 {l_3} 0 
\left\{ 1+(-1)^{k_1+k_2-k} \right\} + b_{l_3}  ~ \cg k 2 {k_1} 1 {k_2} {-1} ~ \cg {l_1} 0 k {-2} {l_3} 2 \left\{ 1+(-1)^{k_1+k_2+l_1+l_3} \right\}   \right] \;,
\ea
\end{widetext}
where $a_{k} = \sqrt{k(k+1)/2}$ and $b_{k} =  \sqrt{(k-1)(k+2)/2}$ .

Explicit evaluation of these terms require direction dependent statistics of the primordial fluctuations and that of the projected lensing potential which in turn would  require specific modeling of non-SI. The formalism presented here is completely generic and hence we refrain from discussing specific models of non-SI. 

\subsubsection*{Reproducing standard lensing results.}
If one assumes that the unlensed temperature field and the projected lensing potential to be SI, then it is expected that we recover the standard weak lensing modifications to the CMB angular power spectra. In the SI case the only BipoSH modes that would contribute are with indices $L' = 0$ and $K =0$ in Equations \ref{beta} and \ref{gamma} and we show that this enforces the second order correction terms to reduce to the standard lensing results with $L=0$ and $l_1 =l_2$ as seen in the following expressions,
\ba
\beta^{LM}_{l_1 l_2} &=& (-1)^{l_1} \Pi_{l_1}\delta_{L 0} \delta_{M 0} \delta_{l_1 l_2 } \sum_{k_1l_3} C^{\psi \psi}_{k_1} C_{l_3} \nn
&\times & \left[ \frac{F(l_1,k_1,l_3)}{\sqrt{4\pi}} \frac{\Pi_{k_1}\Pi_{k_3}}{\Pi_{l_1}} \cg {l_1} 0 {k_1} 0 {l_3} 0 \right]^2 
\;, \label{isolens1} \\
\gamma^{LM}_{l_1 l_2} &=& (-1)^{l_1+1} \; \Pi_{l_1} \delta_{L 0} \delta_{M 0} \delta_{l_1 l_2 } \,\, l_1(l_1+1)C_{l_1} \nn &\times & \left[ \sum_{k_1} \frac{k_1(k_1+1)(2k_1+1)}{8\pi} C^{\psi \psi}_{k_1} \right] 
\;.
\label{isolens2}
\ea
The delta functions in the above equations reduces Eq.~\ref{lens-biposh} to the SI lensed angular power spectrum for which the only 
non-vanishing BipoSH coefficients are of the form $A^{00}_{l_1 l_1} = (-1)^{l_1} \Pi_{l_1} C_{l_1}$. The weak lensing correction to these terms is only due to the terms which are second order in the lensing potential, i.e. $\beta^{00}_{l_1 l_1}$ and $\gamma^{00}_{l_1 l_1}$. The first order correction term does not contribute to the lensed angular power spectrum when the unlensed temperature field is considered to be SI. 
Hence using the correction terms given in Eq.~\ref{isolens1} \& Eq.~\ref{isolens2} in Eq.~\ref{lens-biposh} yields an expression for the lensed CMB angular power spectrum $\tilde{C}_l$ in terms of the unlensed CMB angular power spectrum $C_l$ and the projected lensing potential power spectrum $C_l^{\psi\psi}$,
\ba
\tilde{C}_{l_1}  &=& C_{l_1} - R l_1(l_1+1)C_{l_1} + \sum_{k_1 l_3} C^{\psi \psi}_{k_1} C_{l_3}  \nn &\times & \frac{F(l_3,k_1,l_1)^2}{4\pi} \left[ \frac{\Pi_{k_1}\Pi_{l_3}} {\Pi_{l_1}} \cg {l_1} 0 {k_1} 0 {l_3} 0 \right]^2 \,,
\label{stand-lens}
\ea
where,
\ba
R =\sum_{k_1} \frac{k_1(k_1+1)(2k_1+1)}{8\pi} C^{\psi \psi}_{k_1} \;.
\ea
This result matches exactly with the expression derived for the lensed CMB temperature power spectrum in existing literature \cite{WH} and 
serves as a consistency check for the formalism discussed in this article. 
\section{Modifications to the CMB angular power spectrum}\label{modcl}
The most general expression for the lensing modification to BipoSH spectra is given in Eq.~\ref{lens-biposh} where the individual correction terms are given in Eq.~\ref{alpha}, \ref{beta} and \ref{gamma}. These lensed BipoSH spectra $\tilde{A}^{LM}_{l_1 l_2}$ arise due to coupling of BipoSH spectra $A^{L'M'}_{l_3 l_4}$ of the non-SI unlensed temperature field  and the BipoSH spectra describing the non-SI lensing potential.  Here we explore the contribution of these non-SI terms to the lensed CMB angular power spectrum. 

It is generally believed that searching for off-diagonal power in the harmonic space covariance matrix is the only way to detect deviations from SI. However weak lensing mediates a coupling of power between the diagonal and off-diagonal components of the covariance matrix. Hence in a non-SI universe there could be modifications to the CMB angular power spectrum due to mixing of power between the diagonal and off-diagonal elements of the covariance matrix. In principle these non-standard deviations could be used to test for deviations from SI.

To quantify the effects of non-SI on the lensed CMB temperature power spectrum we evaluate the lensing modification to the BipoSH spectrum $A^{00}_{ll}$. Evaluation of  Eq.~\ref{lens-biposh} for the BipoSH modes $L=0$ results in the following expression,
\ba
\tilde{A}^{00}_{l_1 l_2} &=& A^{00}_{l_1 l_2}+ \;\al^{00}_{l_1l_2}(\psi_{kn},A^{L'M'}_{l_1 l_2}) \nn
	&+& \beta^{00}_{l_1 l_2}(\Psi^{KN}_{k_1 k_2},A^{L'M'}_{l_3 l_4}) + \gamma^{00}_{l_1 l_2}(\Psi^{KN}_{k_1 k_2},A^{L' M'}_{l_3 l_4}) \,, \nn
\tilde{C}_{l_1} &=&  C_{l_1} + \frac{(-1)^{l_1}}{\Pi_{l_1}} \left[ \al^{00}_{l_1l_1} + \beta^{00}_{l_1l_1} + \gamma^{00}_{l_1l_1}\right] \label{mod_cl}\; .
\ea
We consider three scenarios, first we consider the case where the CMB sky at the surface of last scattering is SI but the lensing potential is non-SI, next we consider the case where the CMB sky at the surface of last scattering is non-SI but the lensing potential is SI and finally we evaluate the case when both the CMB at the surface of last scattering and the lensing potential are non-SI.

In the first case, the unlensed CMB sky is SI implying that only $L'=0$ BipoSH spectra are non-vanishing; however the lensing potential is non-SI which means that all BipoSH spectra $K \in [0,1,2,3 \cdots]$ contribute.  However, using properties of the Clebsch-Gordon coefficients \cite{varsha} it can be shown that setting $L'=0$ results in only $K=0$ BipoSH spectra of the lensing potential contributing in all the terms $\alpha$, $\beta$ and $\gamma$. In particular we find that the contribution from the first order term $\alpha^{00}_{l_1 l_2}$ vanishes while the contribution from the second order terms $\beta^{00}_{l_1 l2}$ and $\gamma^{00}_{l_1 l2}$ reduce to the standard lensing terms as in Eq.~\ref{stand-lens}. 

In the next case we have the lensing potential to be SI implying contributions only from $K=0$ BipoSH spectra and a non-SI underlying unlensed CMB power spectrum which mean that in principle any BipoSH mode $L' \in [0,1,2,3 \cdots]$ could contribute. In this case too, setting $K=0$ and using the properties of the Clebsch-Gordon coefficients, it can be shown that only $L'=0$ mode can contribute. This again results in the correction terms reducing to the standard lensing results. 

Thus we see that the corrections to the power spectrum due to non-SI power vanishes on either the lensing potential being SI or the CMB at surface of last scattering being SI. These two cases are significant as it shows that late time non-SI power generation in the lensing potential, for example from direction dependent dark energy, does not effect the angular power spectrum. Therefore any non-standard lensing induced modification to the angular power spectrum in a non-SI universe can only be at second order in the parameter characterizing the deviation from isotropy. 

Finally we consider the case where the lensing potential as well as the unlensed CMB sky are non-SI. The terms in Eq. \ref{mod_cl} which contribute to non-standard correction to the angular power spectrum $C_l$ are for $L' \neq 0$ and $K \neq 0$; the other possible values reduce the correction terms trivially to the standard lensing results as discussed above. Contributions from each of the correction term is explicitly evaluated and can be expressed as
\begin{widetext}
\begin{subequations}
\ba
\al^{00}_{l_1l_1}  &=& 2 (-1)^{l_1} \Pi_{l_1} \sum^{k \ne 0}_{k n l_3} \psi^*_{kn}  A^{kn}_{l_3 l_1} \frac{F(l_3,k,l_1)}{\sqrt{4\pi}} \frac{\Pi_{l_3}}{\Pi_{l_1}\Pi_{k}} 
\cg k 0 {l_1} 0 {l_3} 0 \,,\label{alpha_cl}\\
\beta^{00}_{l_1l_1}  &=& \sum^{K\ne 0}_{K N k_1 k_2} \Psi^{KN}_{k_1 k_2} \sum_{n_1 n_2} \cg K N {k_1} {n_1} {k_2} {n_2} 
\sum^{L^\prime \ne 0}_{L'M' l_3 l_4} A^{L' M'}_{l_3 l_4} (-1)^{l_1+l_3+l_4} \frac{\Pi_{l_3} \Pi_{l_4} }{\Pi_{l_1}} \frac{F(l_1,k_1,l_3) F(l_1,k_2,l_4)}{\sqrt{4\pi}} \label{beta_cl} \nn
&\times& \cg {k_1} 0 {l_1} 0 {l_3} 0 \cg {k_2} 0 {l_1} 0 {l_4} 0  \sum_{m_1 m_3 m_4} (-1)^{m_1} \cg {L'} {M'} {l_3} {m_3} {l_4} {m_4}
\cg {l_1} {m_1} {k_1} {n_1} {l_3} {m_3} \cg {l_1} {-m_1} {k_2} {n_2} {l_4} {m_4}\,, \\
\gamma^{00}_{l_1 l_1} &=& \sum^{K \ne 0}_{K N k_1 k_2} \Psi^{KN}_{k_1 k_2} \sum_{n_1 n_2} \cg K N {k_1} {n_1} {k_2} {n_2} 
\sum^{L^\prime \ne 0}_{L'M' l_3} A^{L' M'}_{l_3 l_1}  \frac{(-1)^{l_1}}{\Pi_{l_1}} \sum_{m_1 m_3} (-1)^{m_1} \cg {L'} {M'} {l_3} {m_3} {l_1} {-m_1} \nn
&\times& \int d \ncap Y^{:a}_{k_1 n_1}(\ncap) Y^{:b}_{k_2 n_2}(\ncap) Y_{l_3 m_3 \; :ab}(\ncap) \ylm{l_1}{m_1} \,. \label{gamma_cl}
\ea
\end{subequations}
\end{widetext}

Thus the non-standard lensing modifications to the lensed CMB temperature anisotropy angular power spectra arising due to the universe being non-SI only comes about due to the coupling between the off diagonal terms of the covariance matrices of the non-SI lensing field and the non-SI CMB temperature field (i.e. only through coupling between BipoSH modes with $K\ne 0$ and $L^\prime\ne0$).
\section{Lensing corrections to cosmic variance}
In this section, we take a diversion to analyze the effects of the weak lensing correction terms $\alpha$, $\beta$ and $\gamma$ in a SI universe. As argued in section \ref{modcl}, weak lensing in a non-SI universe can leave imprints of SI violation in the CMB angular power spectra (See Eq. \ref{mod_cl}). However, if we were to be sitting in an isotropic universe, then the contribution from these non-standard correction terms would vanish, i.e. $\langle\alpha^{00}_{ll}\rangle=\langle\beta^{00}_{ll}\rangle=\langle\gamma^{00}_{ll}\rangle=0$. Note however that the variance of these terms need not vanish, i.e.~$\langle|\alpha^{00}_{ll}|^2\rangle \ne 0,\langle|\beta^{00}_{ll}|^2\rangle \ne 0, \langle|\gamma^{00}_{ll}|^2\rangle \ne 0$. These properties are similar to that of noise, except that this noise is of cosmic origin.

We evaluate the corrections to cosmic variance arising because of weak lensing in a SI universe. To recap, the weak-lensing contributions to the lensed bipolar coefficients are 
\be
\tilde{A}^{LM}_{l_1 l2} = A^{LM}_{l_1 l2} + \alpha^{LM}_{l_1 l2} + \beta^{LM}_{l_1 l2} + \gamma^{LM}_{l_1 l2} + O(\psi^3,\psi^4) \, ,
\ee
which reduce to the angular power spectrum under SI $\tilde{A}^{00}_{l l}=(-1)^{l} \Pi_l \tilde{C}_l$. For the case of a Gaussian and SI CMB sky, it can be shown that the covariance of the BipoSH coefficients \cite{biposh-stats} is given by the following expression,
\ba 
\langle A^{LM}_{l_1 l_2 } {A^{* L'M'}_{l'_1 l'_2}} \rangle &=& C_{l_1} C_{l_2} [\delta_{l_1 l'_1} \delta_{l_2 l'_2} + (-1)^{l_1+l_2+L} \delta_{l_1 l'_2} \delta_{l_2 l'_1}] \nn 
& \times &\delta_{LL'} \delta_{MM'} + C_{l_1} C_{l'_1} \Pi_{l_1} \Pi_{l'_1} (-1)^{l_1+l'_1} \nn 
&\times &\delta_{l_1 l_2} \delta_{l'_1 l'_2} \delta_{L0} \delta_{M 0} \delta_{L' 0} \delta_{M' 0}\, .
\label{biposhvar}
\ea
The variance of the lensed CMB angular power spectrum to first order terms in $C^{\psi}_l$ is given by 
\ba 
\sigma^2_{\tilde{C}_l} &=& \langle \tilde{C}_l \tilde{C}_l \rangle - {\langle \tilde{C}_l \rangle}^2  
\label{sigma}
\ea
Since the $\alpha^{00}_{ll}$ does not contribute at the power spectrum level but has non-zero contribution to the variance, we can write the individual terms in Equation \ref{sigma} as 
\ba 
\langle \tilde{C}_l \rangle &=& C_l + \frac{(-1)^l}{\Pi_l} \{ \langle \beta^{00}_{ll} \rangle + \langle \gamma^{00}_{ll} \rangle \} \nn
{\langle \tilde{C}_l \rangle}^2 &=&  C^2_l + 2 \frac{(-1)^l}{\Pi_l} (C_l \beta^{00}_{ll} + C_l \gamma^{00}_{ll}) + O[(C^{\psi}_l)^2] \\ 
\langle {\tilde{C}_l}^2 \rangle &=& \langle \{ C_l +\frac{(-1)^l}{\Pi_l} (\alpha^{00}_{ll}+\beta^{00}_{ll}+\gamma^{00}_{ll}) \} ^2 \rangle \nn 
&=& \langle C_l^2 \rangle + 2 \frac{(-1)^l}{\Pi_l} [ \langle C_l  \alpha^{00}_{ll} \rangle + \langle C_l  \beta^{00}_{ll} \rangle + \langle C_l  \gamma^{00}_{ll} \rangle ] \nn
&+& \frac{1}{\Pi^2_l} \langle \alpha^{* 00}_{ll} \alpha^{00}_{ll} \rangle  + O[(C^{\psi}_l)^2] \, .
\label{sigma-contrib}
\ea
We evaluate the corrections to cosmic variance arising due to the term linear in the projected lensing potential. The variance correction due to this term can be shown to be linear in the projected lensing potential power spectra $C_l^{\psi}$ while the correction coming from other terms is quadratic in $C_l^{\psi}$ hence is expected to be a sub-dominant. As seen in Equation \ref{sigma-contrib}, the terms contributing to the correction of the cosmic variance and are linear in $C^{\psi}_l$ are $\langle |\alpha^{00}_{ll}|^2 \rangle $, $\langle A^{* 00}_{ll} \alpha^{00}_{ll} \rangle$, $\langle A^{* 00}_{ll} \beta^{00}_{ll} \rangle$ and $\langle A^{* 00}_{ll} \gamma^{00}_{ll} \rangle$. 

The individual terms are evaluated in Appendix \ref{app:cosvar}. Using these correction terms which are relevant in the case of a SI universe, the total cosmic variance is given by the following expression,
\ba 
\sigma^2_{\tilde{C}_l} &=& \frac{2}{2l+1} \tilde{C}^2_l \nn &+&  \frac{8 C^2_l }{2l+1}  
\sum_k C^{\psi}_k \left\{ \frac{F(l, k, l)}{\sqrt{4\pi}} \Pi_k \cg {l} 0 k 0 {l} 0 \right\}^2 \, .
\ea
\section{Extending the formalism to CMB polarization}
The higher order weak lensing correction terms discussed in Section \ref{tlenbips} will affect the CMB polarization field too. As a preliminary study, we lay down the formalism of CMB polarization BipoSH spectra. Weak lensing modifications to these coefficients would follow the same trend as in the temperature case. 

The CMB polarization field is a spin-2 field and can be decomposed into stokes Q$(\ncap)$ and U$(\ncap)$ fields. Using standard convention of decomposing the polarization field into gradient (E) and curl (B) modes which are co-ordinate independent, we can write
\ba
_{\pm}X(\ncap)&=&Q(\ncap) \pm i U(\ncap) \nn
&=&\sum_{l m} \, _{\pm}X_{lm} \ylms{l}{m}{\pm 2}(\ncap) \nn
&=& \sum_{l m} (E_{lm} \pm i B_{lm}) \ylms{l}{m}{\pm 2}(\ncap) \; ,
\ea
where $\ylms{l}{m}{\pm 2}(\ncap)$ are spin-2 spherical harmonics \cite{goldberg, zal-sel}. $a^{E}_{lm} \equiv E_{lm}$ and $a^{B}_{lm} \equiv B_{lm}$ are the E-mode and B-modes spin spherical harmonic coefficients. The Stokes parameter Q and U are real and hence the complex conjugate of the spin-2 field reduces to $_+ X(\ncap)^* = _-X(\ncap)$. The spin spherical harmonic coefficients can be expressed as
\ba
E_{lm} &=& \frac{_+X_{lm} + _-X_{lm}}{2} \nn
B_{lm} &=& \frac{_+X_{lm} - _-X_{lm}}{2i} \; .
\ea
The complex conjugates of the above coefficients are 
\ba 
_+X^*_{lm} &=& (-1)^m _-X_{l,-m} \nn
E^*_{lm} &=& (-1)^m E_{l,-m} \nn
B^*_{lm} &=& (-1)^m B_{l,-m} \, .
\label{pol-par}
\ea
We briefly layout the formalism of polarization BipoSH coefficients in the next section. In the following sections we discuss only the results of the first order lensing corrections to these coefficients and their contribution to the polarization angular power spectrum. 

\subsection{CMB polarization BipoSH coefficients}
Casting aside the constraint of SI, the most generalized CMB polarization two-point correlation function can be expanded in the BipoSH basis using spin-2 spherical harmonics in the following manner, 
\ba
C(\ncap,\ncap') &=&\langle _{\pm} X(\ncap) _{\pm} X(\ncap') \rangle \nn
&=&\langle _{+} X(\ncap) _{+} X(\ncap') \rangle  + \langle _{+} X(\ncap) _{-} X(\ncap') \rangle + \mathrm{c. c.} \nn
&=& \sum_{l m} (E_{l_1 m_1} + iB_{l_1 m_1}) \ylms{l_1}{m_1}{2} \nn 
&\times & \left[ (E_{l_2 m_2} + i B_{l_2 m_2}) \ylms{l_2}{m_2}{2} \right. \nn 
&+& \left. (E_{l_2 m_2} - i B_{l_2 m_2}) \ylms{l_2}{m_2}{-2} \right]+  \mathrm{c. c.} \, ,
\label{pol-xbasis}
\ea
where c.c. are the complex conjugate terms. However, the two-point correlation function can also be expanded in the BipoSH basis as follows
\ba 
C(\ncap,\ncap') &=& \sum_{LM l_1 l_2}  \, _1 A^{LM}_{l_1 l_2} \{ _2 Y_{l_1}(\ncap) \otimes \, _2 Y_{l_2}(\ncap') \}_{LM}  \nn 
&+&  _2 A^{LM}_{l_1 l_2} \{ _{-2} Y_{l_1}(\ncap) \otimes \, _{-2} Y_{l_2}(\ncap') \}_{LM}  \nn 
&+&  _3 A^{LM}_{l_1 l_2} \{ _2 Y_{l_1}(\ncap) \otimes \, _{-2} Y_{l_2}(\ncap') \}_{LM}  \nn
&+& _4 A^{LM}_{l_1 l_2} \{ _{-2} Y_{l_1}(\ncap) \otimes \, _2 Y_{l_2}(\ncap') \}_{LM}) \; ,
\label{pol-bips-basis}
\ea
where $_n A^{LM}_{l_1 l_2} $ are the BipoSH coefficients for the corresponding BipoSH basis. Comparing Eqs. \ref{pol-xbasis}  and \ref{pol-bips-basis} and parity properties of the coefficients as in Eq. \ref{pol-par}, it can easily be shown that ${_1}A^{LM}_{l_1 l_2} ={_2}A^{LM}_{l_1 l_2}$ and ${_3}A^{LM}_{l_1 l_2}  = {_4} A^{LM}_{l_1 l_2}$. 

The + and the - signs on the spin-BipoSH coefficients denote the angular averages of like and unlike terms, viz. ($\langle _+X _+X \rangle$, $\langle _-X _-X \rangle$) and ($\langle _+X _-X \rangle$, $\langle _-X _+X \rangle$) respectively. These coefficients can be expressed in terms of the spin-2 spherical harmonic coefficients in the same way as for the temperature case. 
\ba
_+A^{LM}_{l_1 l_2} &=& \sum_{m_1 m_2} \{ \langle _+X_{l_1 m_1} \; _+X_{l_2 m_2} \rangle + \langle _-X_{l_1 m_1} \; _-X_{l_2 m_2} \rangle\} \nn &\times & \cg L M {l_1} {m_1} {l_2} {m_2} \nn
&=&\sum_{m_1 m_2} \langle (E_{l_1 m_1} E_{l_2 m_2} - B_{l_1 m_1} B_{l_2 m_2}) \cg L M {l_1} {m_1} {l_2} {m_2} \nn 
&=& E^{LM}_{l_1 l_2} - B^{LM}_{l_1 l_2} \; ,
\ea
where $E^{LM}_{l_1 l_2}$ and $B^{LM}_{l_1 l_2}$ are the E-mode and B-mode BipoSH coefficients respectively. Similarly, 
\ba
_-A^{LM}_{l_1 l_2} &=& \sum_{m_1 m_2} \{ \langle _+X_{l_1 m_1} \; _-X_{l_2 m_2} \rangle \langle _-X_{l_1 m_1} \; _+X_{l_2 m_2} \rangle\} 
\nn &\times & \cg L M {l_1} {m_1} {l_2} {m_2} \nn 
&=& E^{LM}_{l_1 l_2} + B^{LM}_{l_1 l_2} \; .
\ea
Under the assumption of SI, the spin-2 BipoSH coefficients reduce to combinations of the standard E-mode and B-mode angular power spectrum as 
\ba
_+A^{00}_{l_1 l_1} &=& (-1)^{l_1} \Pi_{l_1} (C^{EE}_{l_1} - C^{BB}_{l_1}) \nn
_-A^{00}_{l_1 l_1} &=& (-1)^{l_1} \Pi_{l_1} (C^{EE}_{l_1} + C^{BB}_{l_1}) \; .
\ea

\section{Discussions}
The BipoSH basis forms a natural basis to study the CMB two point correlation function of the CMB sky. The angular power spectrum is a subset of the BipoSH spectra. While a SI Gaussian CMB sky is fully characterized by the angular power spectrum, a non-SI CMB sky is characterized by the whole set of BipoSH spectra.  In a SI universe, weak lensing due to LSS surrounding the observer distorts the CMB sky and induces modification to the CMB angular power spectrum which can be evaluated given the angular power spectrum of the projected lensing potential $\psi$.  In this article we study lensing in a non-SI universe, in which the statistics of the CMB sky and the projected lensing potential are specified completely by the BipoSH spectra. 

Analogous to the lensing modifications to the CMB angular power spectrum in a SI universe, we evaluate the lensing modification to the BipoSH spectra due to weak lensing in an non-SI universe. We provide an equation for the BipoSH coefficients of the lensed CMB sky evaluated in terms of the BipoSH spectra of the unlensed CMB sky and those of the projected lensing potential. We study the result under various approximations. 
Firstly we note that in the case of SI CMB sky, lensing of the CMB by the particular realization of LSS surrounding the observer generates BipoSH spectra. We further show that the bipolar map and the bipolar power spectrum proposed as blind tests of SI violation relate to the projected lensing potential and it power spectrum respectively. We note that these estimator are closely related to the optimal lensing estimators \cite{okamoto-hu}.

Next we evaluate the corrections to the CMB angular power spectrum due to the coupling between the non-SI power in the projected lensing potential and the primordial CMB sky.  It is found that this non-standard correction vanishes if either the projected lensing potential or the CMB at the surface of last scattering is SI. This implies that any correction to the power spectrum is at second order in the parameter characterizing the isotropy violation. This analysis also leads to conclusion that any late time isotropy violating phenomena e.g direction dependent expansion, which only makes the lensing map non-SI cannot result in any corrections to the angular power spectrum of the observed sky. 

We also provide an expression describing these non-standard corrections to the angular power spectrum in a non-SI universe. A search for these non-standard corrections in the measured power spectrum can in principle be used to indicate deviations from SI. For instance, the observed temperature angular power spectrum from recent measurement has considerable deviations from standard $\Lambda$CDM angular power spectrum at certain multipole moments like $l = 2, 22$ and $40$ which could be potentially indication of deviations from SI.

We also study the effects of the higher order lensing correction terms in a SI universe. While these correction terms does not contribute  at the angular power spectrum level in a SI universe, they introduce some additional terms in the variance. We evaluate this correction to the cosmic variance in a SI universe due to weak lensing effects. 

Finally we discuss the BipoSH formalism for describing CMB polarization in a non-SI universe. The polarization BipoSH basis consists of combinations of spin-Spherical Harmonics and the corresponding BipoSH coefficients are further characterized as E-mode and B-mode BipoSH coefficients. We can further calculate how these coefficients are modified due to weak lensing as in the temperature case and it can be shown that the modifications to the polarization BipoSH coefficients closely resemble the modifications to the temperature BipoSH coefficients. 

\section*{Acknowledgments}
We are grateful to Barun Pal for useful discussions. AR acknowledges the Council of Scientific and Industrial Research (CSIR), India
for financial support (Grant award no. 20-6/2008(II)E.U.-IV). 
\begin{widetext}
\appendix
\section{Calculation of correction terms to temperature BipoSH coefficients due to lensing}\label{app:cov}
The two point correlation function for the lensed temperature field can be expressed in terms of the
unlensed temperature field and the projected lensing potential through Equation \ref{2ptlens}.
Each of the terms in the two point correlation function of the lensed temperature field in Equation \ref{2ptlens}, 
can be expressed in the harmonic space as
\ba
      \sum_{\substack{L M \\ l_1 l_2}} \tilde{A}^{LM}_{l_1 l_2} \{ Y_{l_1}(\ncap_1) &\otimes& Y_{l_2}(\ncap_2)\}_{LM} 
      = \sum_{L M l_1 l_2} A^{LM}_{l_1 l_2} \{ Y_{l_1}(\ncap_1) \otimes Y_{l_2}(\ncap_2)\}_{LM} \nn 
      &+& \sum_{kn} \psi_{kn} Y^{:a}_{kn}(\ncap_1) \sum_{L' M' l_3 l_4} A^{L'M'}_{l_3 l_4} 
      \{ Y_{l_3 \; :a}(\ncap_1) \otimes Y_{l_4}(\ncap_2)\}_{L'M'} \nn
      &+& \sum_{kn} \psi_{kn} Y^{:a}_{kn}(\ncap_2) \sum_{L' M' l_3 l_4} A^{L'M'}_{l_3 l_4} 
      \{ Y_{l_3}(\ncap_1) \otimes Y_{l_4 \; :a}(\ncap_2)\}_{L'M'} \nn
      &+& \sum_{\substack{K N \\ k_1 k_2}} \Psi^{KN}_{k_1 k_2} \{ Y^{:a}_{k_1}(\ncap_1) \otimes Y^{:b}_{k_2}(\ncap_2) \}_{KN} 
      \sum_{\substack{L'M' \\ l_3 l_4}} A^{L'M'}_{l_3 l_4} \{ Y_{l_3 :a}(\ncap_1) \otimes Y_{l_4 :b}(\ncap_2) \}_{L'M'} \nn
      &+& \frac{1}{2} \sum_{\substack{K N \\ k_1 k_2}} \Psi^{KN}_{k_1 k_2} \{ Y^{:ab}_{k_1}(\ncap_1) \otimes Y_{k_2}(\ncap_2) \}_{KN} 
      \sum_{\substack{L'M' \\ l_3 l_4}} A^{L'M'}_{l_3 l_4} \{ Y_{l_3 :ab}(\ncap_1) \otimes Y_{l_4}(\ncap_2) \}_{L'M'} \nn
      &+& \frac{1}{2} \sum_{\substack{K N \\ k_1 k_2}} \Psi^{KN}_{k_1 k_2} \{ Y^{:ab}_{k_1}(\ncap_2) \otimes Y_{k_2}(\ncap_1) \}_{KN} 
      \sum_{\substack{L'M' \\ l_3 l_4}} A^{L'M'}_{l_3 l_4} \{ Y_{l_3 :ab}(\ncap_2) \otimes Y_{l_4}(\ncap_1) \}_{L'M'}
\ea
Using the orthonormality relation of the BipoSH
\be
\iint d{\ncap_1} d{\ncap_2} \{ Y_{l_1}(\ncap_1) \otimes  Y_{l_2}(\ncap_2)
\}_{LM} 
\{ Y_{l_3}(\ncap_1) \otimes  Y_{l_4}(\ncap_2) \}_{L'M'}^* = \delta_{l_1
l_3}  \delta_{l_2 l_4}  \delta_{L L'} \delta_{M M'},
\label{ortho}
\ee
the lensed temperature BipoSH coefficients $\tilde{A}^{LM}_{l_1 l_2}$ can be obtained in terms of the BipoSH coefficients 
for the unlensed temperature field $A^{L'M'}_{l_3 l_4}$ and the projected lensing potential $\Psi^{KN}_{k_1 k_2}$ as
\ba
	\tilde{A}^{LM}_{l_1 l_2}&=& A^{LM}_{l_1 l_2}+ \left[ \;_1\al^{LM}_{l_1l_2}(\psi_{kn},A^{L'M'}_{l_1 l_2})+
	  \;_2\al^{LM}_{l_1l_2}(\psi_{lm},A^{L'M'}_{l_1 l_2}) \right]
	+ \beta^{LM}_{l_1 l_2}(\Psi^{KN}_{k_1 k_2},A^{L'M'}_{l_3 l_4}) \nn
       &+& \left[ \;_1\gamma^{LM}_{l_1 l_2}(\Psi^{KN}_{k_1 k_2},A^{L' M'}_{l_3 l_4}) 
	+ \;_1\gamma^{LM}_{l_1 l_2}(\Psi^{KN}_{k_1 k_2},A^{L' M'}_{l_3 l_4}) \right] \nn
\ea
\subsection{Calculation of the correction terms $\al^{LM}_{l_1 l_2}$}\label{app:alpha}
The correction terms $\al_1$ and $\al_2$ are linear in the lensing deflection field and are expressed as
\ba
_1\al^{LM}_{l_1 l_2} &=& \iint d\ncap_1 d\ncap_2 ~ \nabla^a \psi(\ncap_1)
\nabla_a^{\ncap_1} \langle T(\ncap_1)T(\ncap_2) \rangle  \lbrace Y_{l_1}(\ncap_1) \otimes Y_{l_2}(\ncap_2)
\rbrace_{LM}^* \nn
&=& \iint d\ncap_1 d\ncap_2 ~  \sum_{kn} \psi_{kn} \ylm{k}{n}^{:a}(\ncap_1)
\sum_{L' M' l_3 l_4} A^{L' M'}_{l_3 l_4} \sum_{m_3 m_4} \cg {L'}{M'}
{l_3}{m_3}{l_4}{m_4} \ylm{l_3}{m_3:a}(\ncap_1) \ylm{l_4}{m_4}(\ncap_2) \nn
&\times & \sum_{m_1 m_2} \cg {L}{M} {l_1}{m_1}{l_2}{m_2}
\ylm{l_1}{m_1}^*(\ncap_1) \ylm{l_2}{m_2}^*(\ncap_2) \nn
&=& \sum_{kn} \psi_{kn} \sum_{L' M' l_3 l_4} A^{L' M'}_{l_3 l_4} \sum_{m_1 m_2 m_3 m_4}  
\cg {L}{M} {l_1}{m_1}{l_2}{m_2} \cg {L'}{M'} {l_3}{m_3}{l_4}{m_4}
\int d\ncap_1 ~\ylm{k}{n}^{:a} \ylm{l_3}{m_3:a} \ylm{l_1}{m_1}^* \nn 
&\times & \int d \ncap_2~ \ylm{l_4}{m_4} \ylm{l_2}{m_2}^* \nn
&=& \sum_{kn} \psi_{kn} \sum_{L' M' l_3} A^{L' M'}_{l_3 l_2} \sum_{m_1 m_2 m_3}  
\cg {L}{M} {l_1}{m_1}{l_2}{m_2} \cg {L'}{M'} {l_3}{m_3}{l_2}{m_2}
\int d\ncap_1 ~\ylm{k}{n}^{:a} \ylm{l_3}{m_3:a} \ylm{l_1}{m_1}^*
\label{b-term}
\ea
\ba
_1\al^{LM}_{l_1 l_2} &=& \sum_{kn} \psi_{kn}  \sum_{L' M' l_3}  A^{L' M'}_{l_3 l_2} \sum_{m_1 m_2 m_3} \cg {L'}{M'} {l_3}{m_3}{l_2}{m_2}
\cg {L}{M} {l_1}{m_1}{l_2}{m_2} I^{m_1 n m_3}_{l_1 k l_3} \;.
\ea
Similarly
\ba
_2\al^{LM}_{l_1 l_2} &=& \sum_{kn}  \psi_{kn} \sum_{L' M' l_3} A^{L' M'}_{l_3 l_1} \sum_{m_1 m_2 m_3} \cg {L'}{M'} {l_3}{m_3}{l_1}{m_1}
\cg {L}{M} {l_1}{m_1}{l_2}{m_2} I^{m_2 n m_3}_{l_2 k l_3} \;.
\ea
\ba
\al^{LM}_{l_1 l_2} = \;_1\al^{LM}_{l_1 l_2} + \;_2\al^{LM}_{l_1 l_2}
\ea
\subsection{Calculation of the correction terms $\beta^{LM}_{l_1 l_2}$}\label{app:beta}
The second order correction term $\beta^{LM}_{l_1 l_2}$ is quadratic in the lensing deflection field and is expressed in terms of the 
BipoSH coefficients of the lensing deflection field, 
\ba
\beta^{LM}_{l_1 l_2} &=& \iint d\ncap_1 d\ncap_2 [\nabla^a_{\ncap_1}
\nabla^b_{\ncap_2} \langle \psi(\ncap_1) \psi(\ncap_2) \rangle]  ~ [\nabla_a^{\ncap_1}
\nabla_b^{\ncap_2} \langle T(\ncap_1) T(\ncap_2) \rangle] ~ \lbrace Y_{l_1}(\ncap_1) \otimes Y_{l_2}(\ncap_2)
\rbrace_{LM}^* \nn
&=& \iint d\ncap_1 d\ncap_2 ~ \left[\nabla^a_{\ncap_1} \nabla^b_{\ncap_2} 
\sum_{KN k_1 k_2} \Psi^{KN}_{k_1 k_2} \sum_{n_1 n_2} \cg {K}{N} {k_1}{n_1}{k_2}{n_2}
\ylm{k_1}{n_1}(\ncap_1) \ylm{k_2}{n_2}(\ncap_2) \right] \nn 
& \times & \left[\nabla_a^{\ncap_1} \nabla_b^{\ncap_2} \sum_{L' M' l_3 l_4} A^{L' M'}_{l_3 l_4}
\sum_{m_3 m_4} \cg {L'}{M'} {l_3}{m_3}{l_4}{m_4} \ylm{l_3}{m_3}(\ncap_1)
\ylm{l_4}{m_4}(\ncap_2) \right]
\sum_{m_1 m_2} \cg {L}{M} {l_1}{m_1}{l_2}{m_2} \ylm{l_1}{m_2}^*(\ncap_1)
\ylm{l_2}{m_2}^*(\ncap_2) \nn
&=& \sum_{K N k_1 k_2} \Psi^{KN}_{k_1 k_2} \sum_{n_1 n_2} \cg {K}{N}
{k_1}{n_1}{k_2}{n_2} \sum_{L' M' l_3 l_4} A^{L' M'}_{l_3 l_4}  \sum_{m_1 m_2 m_3 m_4}
\cg {L}{M} {l_1}{m_1}{l_2}{m_2} ~ \cg {L'}{M'} {l_3}{m_3}{l_4}{m_4} ~ I^{m_1 n_1 m_3}_{l_1 k_1 l_3} ~ I^{m_2 n_2 m_4}_{l_2 k_2 l_4} \, . 
\label{c-term}
\ea
\subsection{Calculation of the correction terms $\gamma^{LM}_{l_1 l_2}$} \label{app:gamma}
The two components of the other second order correction term $\gamma^{LM}_{l_1 l_2}$ are also expressed in terms of the 
BipoSH coefficients of the lensing deflection field, 
\ba
_1\gamma^{LM}_{l_1 l_2} &=& \frac{1}{2} \iint d\ncap_1 ~ d\ncap_2 ~ [\nabla^a_{\ncap_1} \nabla^b_{\ncap_1} \langle \psi(\ncap_1)\psi(\ncap_1) \rangle] 
	~ [\nabla_a^{\ncap_1} \nabla_b^{\ncap_1} \langle T(\ncap_1) T(\ncap_2) \rangle] ~ \lbrace Y_{l_1}(\ncap_1) \otimes Y_{l_2}(\ncap_2)
\rbrace_{LM}^* \nn
&=& \frac{1}{2} \iint d\ncap_1 ~ d\ncap_2 ~ \left[\nabla^a_{\ncap_1} \nabla^b_{\ncap_1} \sum_{K N k_1 k_2} \Psi^{KN}_{k_1 k_2} 
\sum_{n_1 n_2} \cg {K}{N} {k_1}{n_1}{k_2}{n_2} \ylm{k_1}{n_1}(\ncap_1) \ylm{k_2}{n_2}(\ncap_1) \right]\nn 
& \times & \left[\nabla_a^{\ncap_1} \nabla_b^{\ncap_1} \sum_{L' M' l_3 l_4} A^{L' M'}_{l_3 l_4}
\sum_{m_3 m_4} \cg {L'}{M'} {l_3}{m_3}{l_4}{m_4} \ylm{l_3}{m_3}(\ncap_1) \ylm{l_4}{m_4}(\ncap_2) \right]
\sum_{m_1 m_2} \cg {L}{M} {l_1}{m_1}{l_2}{m_2} \ylm{l_1}{m_1}^*(\ncap_1) \ylm{l_2}{m_2}^*(\ncap_2) \nn
&=& \frac{1}{2} \sum_{KN k_1 k_2} \Psi^{KN}_{k_1 k_2} \sum_{n_1 n_2} \cg {K}{N} {k_1}{n_1}{k_2}{n_2}
\sum_{L' M' l_3 l_4} A^{L' M'}_{l_3 l_4} \sum_{m_1 m_2 m_3 m_4} \cg {L}{M} {l_1}{m_1}{l_2}{m_2} ~ \cg {L'}{M'} {l_3}{m_3}{l_4}{m_4} \nn
&\times& \int d\ncap_1 \ylm{k_1}{n_1}^{:a}(\ncap_1) \ylm{k_2}{n_2}^{:b}(\ncap_1) \ylm{l_3}{m_3 \; :ab}(\ncap_1) \ylm{l_1}{m_1}^*(\ncap_1) 
\int d\ncap_2 \ylm{l_4}{m_4}(\ncap_2) \ylm{l_2}{m_2}^*(\ncap_2) \nn
&=& \frac{1}{2}\sum_{K N k_1 k_2} \Psi^{KN}_{k_1 k_2} \sum_{n_1 n_2} \cg {K} {N} {k_1} {n_1} {k_2} {n_2} 
\sum_{L' M' l_3} A^{L' M'}_{l_3 l_2} \sum_{m_1 m_2 m_3}  \cg {L} {M} {l_1} {m_1} {l_2} {m_2} \cg {L'} {M'} {l_3} {m_3} {l_2} {m_2} 
J^{m_1 n_1 m_3 n_2}_{l_1 k_1 l_3 k_2} .
\ea
Similarly, 
\ba
_2\gamma^{LM}_{l_1 l_2} &=& \frac{1}{2}\sum_{K N k_1 k_2} \Psi^{KN}_{k_1 k_2} \sum_{n_1 n_2} \cg {K} {N} {k_1} {n_1} {k_2} {n_2} 
\sum_{L' M' l_3} A^{L' M'}_{l_3 l_1} \sum_{m_1 m_2 m_3}  \cg {L} {M} {l_1} {m_1} {l_2} {m_2} \cg {L'} {M'} {l_3} {m_3} {l_1} {m_1} 
J^{m_2 n_1 m_3 n_2}_{l_2 k_1 l_3 k_2}.
\ea
\be
\gamma^{LM}_{l_1 l_2} = \;_1\gamma^{LM}_{l_1 l_2} + \;_2\gamma^{LM}_{l_1 l_2} \; .
\label{d-term}
\ee
\section{Evaluating the integrals}\label{app:int}
The integral appearing in equation \ref{i-int} can be evaluated by performing integration by parts and the using the identity 
$\nabla^2 \ylm{l}{m}(\ncap) = -l(l+1)\ylm{l}{m}(\ncap)$ as in \cite{DG-DS}, 
\ba
I^{m m_1 m_2}_{l l_1 l_2} &=& \int d{\ncap }\ylm{l}{m}^*(\ncap) \ylm{l_1}{m_1}^{:a}(\ncap) \ylm{l_2}{m_2 \: :a}(\ncap)) \nn
&=& \frac{1}{2} \left[ l_1(l_1+1) + l_2(l_2+1) - l(l+1)\right] \int d{\ncap} \ylm{l}{m}^*(\ncap) \ylm{l_1}{m_1}(\ncap) \ylm{l_2}{m_2}(\ncap)) \nn
&=& \frac{1}{2} \left[ l_1(l_1+1) + l_2(l_2+1) - l(l+1)\right]  
\frac{\Pi_{l_1 l_2}}{\sqrt{4\pi} \Pi_{l}} \cg {l} {0} {l_1} {0} {l_2} {0} \cg {l} {m} {l_1} {m_1} {l_2} {m_2}.
\ea
The second order correction term $\gamma^{LM}_{l_1 l_2}$, contains an integral $J^{m_1 n_1 m_3 n_2}_{l_1 k_1 l_3 k_2}$ as in equation \ref{j-int} where
\ba
J^{m_1 n_1 m_3 n_2}_{l_1 k_1 l_3 k_2} = 
\int  d{\ncap_1} \ylm{l_1}{m_1}^*(\ncap_1) \ylm{k_1}{n_1}^{:a}(\ncap_1) \ylm{k_2}{n_2}^{:b}(\ncap_1) \ylm{l_3}{m_3 \: :ab}(\ncap_1) \, .
\ea
This integral can be expressed in terms of spin spherical harmonics $_sY_{lm}$. We choose a convenient basis to study spin spherical harmonics 
\be
\mathbf{m}_{\pm}= \frac{1}{\sqrt{2}} (\hat{e}_\theta \mp i\hat{e}_{\phi}).
\ee
It is seen that $m_+$ and $m_-$ are complex conjugate of each other i.e. $m^*_{\pm}=m_{\mp}$ and act as the lowering and raising operators 
respectively \cite{goldberg},
\ba
\mathbf{m}_+ \cdot \nabla \ylms{l}{m}{s} = -
\sqrt{\frac{(l+s)(l-s+1)}{2}} \: \ylms{l}{m}{s-1} \nn
\mathbf{m}_- \cdot \nabla \ylms{l}{m}{s} =  
\sqrt{\frac{(l-s)(l+s+1)}{2}} \: \ylms{l}{m}{s+1} .
\ea
In this basis, the derivative of a spherical harmonic is given as
\ba
\nabla \ylm{l}{m} = \sqrt{\frac{l(l+1)}{2}} \left[ \mathbf{m}_+ \: 
\ylms{l}{m}{1} - \mathbf{m}_- \: \ylms{l}{m}{-1} \right] \; . 
\label{grad-ylm}
\ea
Using these notations, the $\mathcal{J}$-integral can be expressed as
\ba
J^{m_1 n_1 m_3 n_2}_{l_1 k_1 l_3 k_2} &=& a_{k_1} a_{k_2}
\int  d{\ncap} ~ \ylm{l_1}{m_1}^* \left[ \mathbf{m}_{+a} \;_1Y_{k_1 n_1} - \mathbf{m}_{-a} \;_{-1}Y_{k_1 n_1} \right]
\left[ \mathbf{m}_{+b} \;_1Y_{k_2 n_2} - \mathbf{m}_{-b} \;_{-1}Y_{k_2 n_2} \right] \ylm{l_3}{m_3 \: :ab} \nn
&=& a_{k_1} a_{k_2} \int  d{\ncap} ~ \ylm{l_1}{m_1}^* \left[ \ylms{k_1}{n_1}{1}  \ylms{k_2}{n_2}{1} \mathbf{m}_{+a}\mathbf{m}_{+b} + 
\ylms{k_1}{n_1}{-1}  \ylms{k_2}{n_2}{-1} \mathbf{m}_{-a}\mathbf{m}_{-b} \right. \nn 
&-& \left.  \ylms{k_1}{n_1}{1}  \ylms{k_2}{n_2}{-1} \mathbf{m}_{+a}\mathbf{m}_{-b} - 
\ylms{k_1}{n_1}{-1}  \ylms{k_2}{n_2}{1} \mathbf{m}_{-a}\mathbf{m}_{+b} \right] \ylm{l_3}{m_3 \: :ab} \; ,
\ea
noting that $a_{k} = \sqrt{k(k+1)/2}$ . To evaluate this integral, we utilize identities for spin spherical harmonics 
from \cite{WH} and \cite{WH-MW}
\ba
\ylms{k_1}{n_1}{s_1}(\ncap) \ylms{k_2}{n_2}{s_2}(\ncap) &=& \sum_{kns} \frac{\Pi_{k_1k_2}}{\sqrt{4\pi}\Pi_k} 
\cg k {-s} {k_1} {-s_1} {k_2} {-s_2} \cg k n {k_1} {n_1} {k_2} {n_2} \ylms{k}{n}{s}(\ncap) \nn 
\int d{\ncap} \ylms{l_1}{m_1}{s_1}^* \ylms{l_2}{m_2}{s_2} \ylms{l_3}{m_3}{s_3} 
&=& \frac{\Pi_{l_2}\Pi_{l_3}}{\sqrt{4\pi} \Pi_{l_1}} \cg {l_1} {-s_1} {l_2} {-s_2} {l_3} {-s_3} \cg {l_1} {m_1} {l_2} {m_2} {l_3} {m_3} \; .
\ea
The first term of the $\mathcal{J}$-integral simplifies to
\ba
\mathcal{J}_1 &=& a_{k_1} a_{k_2} \int  d{\ncap} ~ \ylm{l_1}{m_1}^* \ylms{k_1}{n_1}{1}  \ylms{k_2}{n_2}{1} (\mathbf{m}_{+a} \nabla^a) 
(\mathbf{m}_{+b} \nabla^b) \ylm{l_3}{m_3} \nn
&=& a_{k_1} a_{k_2} \sum_{kns} \frac{\Pi_{k_1k_2}}{\sqrt{4\pi}\Pi_k} \cg k {-s} {k_1} {-1} {k_2} {-1} \cg k n {k_1} {n_1} {k_2} {n_2}
\int  d{\ncap} ~ \ylm{l_1}{m_1}^* \ylms{k}{n}{s} (\mathbf{m}_{+a} \nabla^a)  \times (-a_{l_3}) \ylms{l_3}{m_3}{-1} \nn
&=& - a_{k_1} a_{k_2} a_{l_3} \sum_{kn} \frac{\Pi_{k_1k_2}}{\sqrt{4\pi}\Pi_k} \cg k {-2} {k_1} {-1} {k_2} {-1} \cg k n {k_1} {n_1} {k_2} {n_2}
\int  d{\ncap} ~ \ylm{l_1}{m_1}^* \ylms{k}{n}{2} \times (-b_{l_3}) \ylms{l_3}{m_3}{-2} \nn
&=& a_{k_1} a_{k_2} a_{l_3} b_{l_3} \sum_{kn} \frac{\Pi_{k_1k_2}}{\sqrt{4\pi}\Pi_k} \cg k {-2} {k_1} {-1} {k_2} {-1} \cg k n {k_1} {n_1} {k_2} {n_2}
\frac{\Pi_{k}\Pi_{l_3}}{\sqrt{4\pi} \Pi_{l_1}} \cg {l_1} {0} {k} {-2} {l_3} {2} \cg {l_1} {m_1} {k} {n} {l_3} {m_3} \nn
&=& \frac{a_{k_1} a_{k_2} a_{l_3} b_{l_3} }{4\pi}\frac{\Pi_{k_1} \Pi_{k_2} \Pi_{l_3}}{\Pi_{l_1}}  \sum_{k}  \cg k {-2} {k_1} {-1} {k_2} {-1}
\cg {l_1} {0} {k} {-2} {l_3} {2} \sum_{n}  \cg k n {k_1} {n_1} {k_2} {n_2} \cg {l_1} {m_1} {k} {n} {l_3} {m_3}  \; .
\ea
Since the Clebsch Gordon coefficients $\cg {l} m {l_1} {m_1} {l_2} {m_2}$ are valid only for $m=m_1+m_2$, the sum over $s$ gets fixed to $s=2$ in this case. 
Similarly the other terms in the $\mathcal{J}$ integral are
\ba
\mathcal{J}_2 &=& \frac{a_{k_1} a_{k_2} a_{l_3} b_{l_3} }{4\pi}\frac{\Pi_{k_1} \Pi_{k_2} \Pi_{l_3}}{\Pi_{l_1}}  \sum_{k} \cg k {2} {k_1} {1} {k_2} {1}
\cg {l_1} {0} {k} {2} {l_3} {-2} \sum_{n}  \cg k n {k_1} {n_1} {k_2} {n_2} \cg {l_1} {m_1} {k} {n} {l_3} {m_3}  \nn
\mathcal{J}_3 &=& \frac{a_{k_1} a_{k_2} a_{l_3}^2}{4\pi}\frac{\Pi_{k_1} \Pi_{k_2} \Pi_{l_3}}{\Pi_{l_1}}  \sum_{k} \cg k {0} {k_1} {-1} {k_2} {1}
\cg {l_1} {0} {k} {0} {l_3} {0} \sum_{n}  \cg k n {k_1} {n_1} {k_2} {n_2} \cg {l_1} {m_1} {k} {n} {l_3} {m_3}  \nn
\mathcal{J}_4 &=& \frac{a_{k_1} a_{k_2} a_{l_3}^2}{4\pi}\frac{\Pi_{k_1} \Pi_{k_2} \Pi_{l_3}}{\Pi_{l_1}}  \sum_{k} \cg k {0} {k_1} {1} {k_2} {-1}
\cg {l_1} {0} {k} {0} {l_3} {0} \sum_{n}  \cg k n {k_1} {n_1} {k_2} {n_2} \cg {l_1} {m_1} {k} {n} {l_3} {m_3}  \nn
\ea
Using the identity from \cite{varsha}
\be
\cg l m {l_1} {m_1} {l_2} {m_2} = (-1)^{l_1+l_2 - l} ~ \cg l {-m} {l_1} {-m_1} {l_2} {-m_2} \; ,
\ee
the $\mathcal{J}$-integral in equation \ref{j-simple} simplifies to
\ba
J^{m_1 n_1 m_3 n_2}_{l_1 k_1 l_3 k_2} &=& \frac{a_{k_1} a_{k_2} a_{l_3}}{4\pi}\frac{\Pi_{k_1} \Pi_{k_2} \Pi_{l_3}}{\Pi_{l_1}} \sum_{k} 
\left[ a_{l_3} \cg k {0} {k_1} {1} {k_2} {-1} \cg {l_1} {0} {k} {0} {l_3} {0} \left\{ 1+(-1)^{k_1+k_2-k} \right\} \right. \nn 
&+& \left. b_{l_3} \cg k {2} {k_1} {1} {k_2} {1} \cg {l_1} {0} {k} {2} {l_3} {-2} \left\{ 1+(-1)^{k_1+k_2+l_3-l_1}\right\} \right]
\sum_{n}  \cg k n {k_1} {n_1} {k_2} {n_2} \cg {l_1} {m_1} {k} {n} {l_3} {m_3} 
\ea
If the unlensed CMB temperature field and the lensing deflection field is assumed to be SI, i.e. $L'=M'=0$ and $K=N=0$, 
the second order correction term $\gamma^{LM}_{l_1 l_2}$ in equation \ref{gamma} reduces to 
\ba
\gamma^{LM}_{l_1 l_2} &=& \frac{1}{2} \sum_{k_1 n_1} (-1)^{n_1} C_{k_1}^{\psi \psi} \sum_{m_1 m_2} \cg L M {l_1} {m_1} {l_2} {m_2} 
\left[ (-1)^{m_2} C_{l_2} J^{m_1 n_1 -m_2 -n_1}_{l_1 k_1 l_2 k_1} + (-1)^{m_1} C_{l_1} J^{m_2 n_1 -m_1 -n_1}_{l_2 k_1 l_1 k_1} \right] , 
\ea
where we need to evaluate the quantity, 
\ba
\sum_{n_1} (-1)^{n_1} J^{m_1 n_1 -m_2 -n_1}_{l_1 k_1 l_2 k_1} &=& \frac{a_{k_1} a_{k_1} a_{l_2}}{4\pi}\frac{\Pi_{k_1} \Pi_{k_1} \Pi_{l_2}}{\Pi_{l_1}} 
\sum_{k} 
\ea
the $\mathcal{J}$ integral
\ba
J_1 &=& a_{k_1} a_{k_2} \sum_{kns} 
\frac{\Pi_{k_1 k_2}}{\sqrt{4\pi}\Pi_k} \cg k {-s} {k_1} {-s_1} {k_2} {-s_2} \cg k n {k_1} {n_1} {k_2} {n_2} \nn
&\times& \int d{\ncap} ~ \ylm{l_1}{m_1}^* \ylms{k}{n}{s} 
\left[ \mathbf{m}_{+a} \mathbf{m}_{+b} + \mathbf{m}_{-a}\mathbf{m}_{-b} - 
\mathbf{m}_{+a} \mathbf{m}_{-b} - \mathbf{m}_{-a} \mathbf{m}_{+b} \right] \ylm{l_3}{m_3 \: :ab} \nn
&=& a_{k_1} a_{k_2} \sum_{kn} \frac{\Pi_{k_1 k_2}}{\sqrt{4\pi}\Pi_k} \cg k 0 {k_1} 0 {k_2} 0 \cg k n {k_1} {n_1} {k_2} {n_2} ~ \mathcal{I} .
\label{j-simple}
\ea
The first term in the integral $\mathcal{I}$ is
\ba
\mathcal{I}_1 &=& \int d{\ncap} ~ \ylm{l_1}{m_1}^* \ylm{k}{n} ~ (\mathbf{m}_{+a} \nabla^a) (\mathbf{m}_{+b} \nabla^b \ylm{l_3}{m_3}) \nn
&=& -a_{l_3} \int d{\ncap} ~ \ylm{l_1}{m_1}^* \ylm{k}{n} ~ (\mathbf{m}_{+a} \nabla^a  \ylms{l_3}{m_3}{-1}) \nn
&=& a_{l_3} \sqrt{\frac{(l_3-1)(l_3+2)}{2}} \int d{\ncap} ~ \ylm{l_1}{m_1}^* \ylm{k}{n} \ylms{l_3}{m_3}{-2} = 0 .
\ea
The above term vanishes for the integral of product of two spin-zero and one spin 2 spherical harmonic functions. 
The second term in the integral $\mathcal{I}$ vanish by similar arguments. The third term in the integral turns out to be same as the fourth term 
and is given as 
\ba
\mathcal{I}_3 &=&  - \int d{\ncap} ~ \ylm{l_1}{m_1}^* \ylm{k}{n} ~ (\mathbf{m}_{+a} \nabla^a) (\mathbf{m}_{-b} \nabla^b \ylm{l_3}{m_3}) \nn  
&=& -a_{l_3} \int d{\ncap} ~ \ylm{l_1}{m_1}^* \ylm{k}{n} ~ (\mathbf{m}_{+a} \nabla^a  \ylms{l_3}{m_3}{1}) \nn
&=& a_{l_3}^2 \int d{\ncap} ~ \ylm{l_1}{m_1}^* \ylm{k}{n} ~ \ylm{l_3}{m_3} \; = \mathcal{I}_4 \, .
\label{j-nonzero}
\ea
Using the above equation and the identity from \cite{varsha}
\be
\int d{\ncap} ~ \ylm{l_1}{m_1}^* \ylm{l_2}{m_2} \ylm{l_3}{m_3}  = \frac{\Pi_{l_2}\Pi_{l_3}}{\sqrt{4\pi}\Pi_{l_1}} \cg {l_1} 0 {l_2} 0 {l_3} 0
\cg {l_1} {m_1} {l_2} {m_2} {l_3} {m_3} ,
\ee
the $\mathcal{J}$-integral in equation \ref{j-simple} simplifies to
\ba
J^{m_1 n_1 m_3 n_2}_{l_1 k_1 l_3 k_2} &=& 2 a_{k_1} a_{k_2} a_{l_3}^2 \sum_{kn} \frac{1}{4\pi}
\frac{\Pi_{k_1} \Pi_{k_2} \Pi_{l_3} }{\Pi_{l_1}} \cg k 0 {k_1} 0 {k_2} 0 \cg k n {k_1} {n_1} {k_2} {n_2} \cg {l_1} 0 {k} 0 {l_3} 0
\cg {l_1} {m_1} {k} {n} {l_3} {m_3} \; . 
\ea
\section{Calculation of the correction terms in the cosmic variance \label{app:cosvar}}
The variance of the lensed CMB angular power spectrum to first order terms in $C^{\psi}_l$ is given by 
\ba 
\sigma^2_{\tilde{C}_l} &=& \langle \tilde{C}_l \tilde{C}_l \rangle - {\langle \tilde{C}_l \rangle}^2  \nn 
&=& \langle C_l^2 \rangle + \frac{2}{\Pi^2_l} [ \langle A^{*00}_{ll}  \alpha^{00}_{ll} \rangle + \langle  A^{*00}_{ll}   \beta^{00}_{ll} \rangle + \langle  A^{*00}_{ll}   \gamma^{00}_{ll} \rangle ] + \frac{1}{\Pi^2_l} \langle \alpha^{* 00}_{ll} \alpha^{00}_{ll} \rangle \nn 
&-& \left[ {\langle C_l \rangle}^2 + \frac{2}{\Pi^2_l} \left[ \langle  A^{*00}_{ll}  \rangle \langle \beta^{00}_{ll} \rangle + \langle  A^{*00}_{ll}  \rangle \langle \gamma^{00}_{ll} \rangle \right] \right] \, ,
\label{sigma-mod}
\ea
where we have used the relation $ A^{00}_{ll} = (-1)^l \Pi_l C_l$ and the correction terms $\alpha, \beta$ and $\gamma$ are given in Equations \ref{alpha_cl}, \ref{beta_cl} and \ref{gamma_cl}. Firstly, we focus on the terms contributing to ${\langle \tilde{C}_l \rangle}^2$ and find 
\ba 
\langle A^{*00}_{ll} \rangle \langle \beta^{00}_{ll} \rangle &=&(-1)^l \Pi_l C_l (-1)^l \Pi_l \sum_{k l_1} C_{l_1} C^{\psi}_k {\left[ \frac{F(l_1 k l)}{\sqrt{4\pi}} \frac{\Pi_k \Pi_{l_1}}{\Pi_l} \cg l 0 k 0 {l_1} 0 \right] }^2 \nn 
&=& C_l \sum_{k l_1} C_{l_1} C^{\psi}_k {\left[ \frac{F(l_1 k l)}{\sqrt{4\pi}} \Pi_k \Pi_{l_1} \cg l 0 k 0 {l_1} 0 \right] }^2 \nn
\langle A^{* 00}_{ll} \rangle \langle \gamma^{00}_{ll} \rangle &=& (-1)^l \Pi_l C_l (-1)^{l+1} \Pi_l l(l+1)C_l R 
= - l(l+1) R C^2_l \Pi^2_l  \, .
\label{betagamma0}
\ea
Next we evaluate the terms which contribute to $\langle \tilde{C}^2_l \rangle$ in Equation \ref{sigma-mod}. The term $\langle A^{* 00}_{ll} \alpha^{00}_{ll} \rangle$ reduces to zero as $\alpha^{00}_{ll}$ does not contribute at power spectrum level. The terms which contribute are 
the cross-terms $\langle A^{* 00}_{ll} \beta^{00}_{ll} \rangle$, $\langle A^{* 00}_{ll} \gamma^{00}_{ll} \rangle$ and the covariance term 
$\langle |\alpha^{00}_{ll}|^2 \rangle $. Using the covariance relation for bipolar coefficients from Equation \ref{biposhvar}, we find
\be 
\langle A^{* 00}_{ll} A^{L M}_{l_1 l_2} \rangle = C^2_l \left( \delta_{l l_1} \delta_{l l_2} + (-1)^{l+l+0} \delta_{l l_2} \delta_{l l_1} \right) \delta_{L 0} \delta_{M 0} + C_l C_{l_1} \Pi_{l} \Pi_{l_1} (-1)^{l+l_1} \delta_{l_1 l_2} \delta_{L 0} 
\label{biposhvar-simpl}
\ee
Using the above Equation \ref{biposhvar-simpl}, the first cross-term is 
\ba 
\langle A^{* 00}_{ll} \beta^{00}_{ll} \rangle &=& \sum_{k n l_1} (-1)^{n} C^{\psi}_k \left( 2 C^2_l \delta_{l l_1} \delta_{l l_2} + C_l C_{l_1} \Pi_l \Pi_{l_1} (-1)^{l+l_1} \delta_{l_1 l_2} \right) \sum_{m m_1} \frac{(-1)^{l_1-m_1}}{\Pi_{l_1}} \frac{(-1)^{l-m}}{\Pi_l} I^{m n m_1}_{l k l_1} I^{-m -n -m_1}_{l k l_1} \nn 
&=& 2 \sum_k C^{\psi}_k C^2_l {\left[ \frac{F(l k l)}{\sqrt{4\pi}} \Pi_k \cg l 0 k 0 l 0 \right] }^2 \frac{1}{\Pi^2_l} \sum_m \sum_{n m_1} (\cg l m k n l {m_1})^2 + \sum_{k l_1} C^{\psi}_k  C_l C_{l_1} {\left[ \frac{F(l_1 k l)}{\sqrt{4\pi}} \Pi_k \Pi_{l_1} \cg l 0 k 0 {l_1} 0 \right] }^2 \frac{1}{\Pi^2_l} \sum_m \sum_{n m_1} (\cg l m k n {l_1} {m_1})^2 \nn 
&=& \langle A^{00}_{ll} \rangle \langle \beta^{00}_{ll} \rangle  + 2 \sum_{k} C^{\psi}_k C^2_l {\left[ \frac{F(l k l)}{\sqrt{4\pi}} \Pi_k 
\cg l 0 k 0 l 0 \right] }^2 \, ,
\label{beta1}
\ea
where we have used the relation for sum of products of two Clebsch-Gordon coefficients from \cite{varsha}
\be 
\sum_{m_1 m_3} \cg {l_3} {m_3} {l_1} {m_1} {l_2} {m_2} \cg {l_3} {m_3} {l_1} {m_1} {l_4} {m_4} = \frac{\Pi^2_{l_3}}{\Pi^2_{l_2}} \delta_{l_2 l_4} \delta_{m_2 m_4} \, .
\ee
and the relation $\sum_m  = \Pi^2_l$. The second cross-term can be evaluated as follows
\ba 
\langle A^{* 00}_{ll} \gamma^{00}_{ll} \rangle &=& \sum_{k n m} \frac{(-1)^n}{\Pi^2_l} C^{\psi}_k (2C^2_l + \Pi^2_l C^2_l) J^{m n -m -n}_{l k l k} = -l(l+1) \{2C^2_l + C^2_l\Pi^2_l\} R \nn 
&=& \langle A^{* 00}_{ll} \rangle \langle \gamma^{00}_{ll} \rangle - 2 C^2_l l(l+1)R \, .
\label{gamma1}
\ea
The covariance of the $\alpha$ term is given by 
\ba 
\langle  \alpha^{00}_{ll}   \alpha^{* 00}_{ll} \rangle &=& 4 \Pi^2_l \sum_{k n l_1} \sum_{k' n' l'_1} \langle \psi^*_{kn} \psi_{k' n'} \rangle 
\langle A^{kn}_{l_1 l} A^{* k' n'}_{l'_1 l} \rangle \frac{F(l_1 k l)F(l'_1 k l)}{4\pi} \frac{\Pi_{l_1}\Pi_{l'_1}}{\Pi^2_l \Pi_k \Pi_{k'}} \cg k 0 l 0 {l_1} 0 \cg {k'} 0 l 0 {l'_1} 0 \nn
&=& 4 \sum_{k n l_1} \sum_{k' n' l'_1} C^{\psi}_k \delta_{k k'} \delta_{n n'} \left[ C_{l_1} C_l \{ \delta_{l_1 l'_1} + (-1)^{l_1+l+k} \delta_{l_1 l} \delta_{l'_1 l} \} \right] \frac{F(l_1 k l)F(l'_1 k l)}{4\pi} \frac{\Pi_{l_1}\Pi_{l'_1}}{\Pi_k \Pi_{k'}} \cg k 0 l 0 {l_1} 0 \cg {k'} 0 l 0 {l'_1} 0  \nn 
&=& 4 \sum_{k l_1} C^{\psi}_k C_{l_1} C_l {\left[\frac{F(l_1 k l)}{\sqrt{4\pi}} \frac{\Pi_{l_1}\Pi_k}{\Pi_{l}} \cg l 0 k 0 {l_1} 0 \right]}^2
+ 4 \sum_{k} C^{\psi}_k C^2_{l} {\left[ \frac{F(l k l)}{\sqrt{4\pi}} \Pi_k \cg l 0 k 0 l 0 \right]}^2 \nn 
&=& \frac{4}{\Pi^2_l} \langle A^{00}_{ll} \rangle \langle \beta^{00}_{ll} \rangle + 4 \sum_{k} C^{\psi}_k C^2_{l} {\left[ \frac{F(l k l)}{\sqrt{4\pi}} \Pi_k \cg l 0 k 0 l 0 \right]}^2 \, .
\label{alpha1}
\ea
Note that the cosmic variance for the unlensed CMB power spectrum is given by $\langle C^2_l \rangle - {\langle C_l \rangle}^2= 2C^2_l/(2l+1)$. Also the lensed CMB power spectrum is 
\ba
\tilde{C}_{l}  &=& C_{l} - R l(l+1)C_l + \sum_{k l_1} C^{\psi}_{k} C_{l_1}  \frac{F(l_1,k,l)^2}{4\pi} \left[ \frac{\Pi_{k}\Pi_{l}} {\Pi_{l}} \cg {l} 0 {k} 0 {l_1} 0 \right]^2 \,.
\ea
Retaining terms which are first order in the lensing potential 
\ba
\tilde{C}^2_{l}  &\approx & C^2_{l} -2 R l(l+1)C^2_l + 2 C_l \sum_{k l_1} C^{\psi}_{k} C_{l_1}\frac{F(l_1,k,l)^2}{4\pi} \left[ \frac{\Pi_{k}\Pi_{l}} {\Pi_{l}} \cg {l} 0 {k} 0 {l_1} 0 \right]^2 \,. 
\ea
Using the equations \ref{betagamma0}, \ref{beta1}, \ref{gamma1} and \ref{alpha1}, the cosmic variance for the lensed CMB power spectrum for terms which are linear in $C^{\psi}_l$ can be expressed as follows,
\ba 
\sigma^2_{\tilde{C}_l} &=& \langle C^2_l \rangle + \frac{2}{\Pi^2_l} \left[ \langle A^{00}_{ll} \rangle \langle \beta^{00}_{ll} \rangle  + 2 \sum_{k} C^{\psi}_k C^2_l {\left[ \frac{F(l k l)}{\sqrt{4\pi}} \Pi_k \cg l 0 k 0 l 0 \right] }^2 + \langle A^{* 00}_{ll} \rangle \langle \gamma^{00}_{ll} \rangle - 2 C^2_l l(l+1)R \right] \nn
&+& \frac{1}{\Pi^2_l} \left[ \frac{4}{\Pi^2_l} \langle A^{00}_{ll} \rangle \langle \beta^{00}_{ll} \rangle + 4 \sum_{k} C^{\psi}_k C^2_{l} {\left[ \frac{F(l k l)}{\sqrt{4\pi}} \Pi_k \cg l 0 k 0 l 0 \right]}^2 \right] -  {\langle C_l \rangle}^2 -\frac{2}{\Pi^2_l}  \left[ \langle A^{00}_{ll} \rangle \langle \beta^{00}_{ll} \rangle + \langle A^{00}_{ll} \rangle \langle \gamma^{00}_{ll} \rangle \right] \nn 
&=& \frac{2}{\Pi^2_l} \left[ C^2_l + 2 C^2_l \sum_{k} C^{\psi}_k  {\left[ \frac{F(l k l)}{\sqrt{4\pi}} \Pi_k \cg l 0 k 0 l 0 \right] }^2 - 2 C^2_l l(l+1)R + \frac{2}{\Pi^2_l} \langle A^{00}_{ll} \rangle \langle \beta^{00}_{ll} \rangle + 2 C^2_{l} \sum_{k} C^{\psi}_k {\left[ \frac{F(l k l)}{\sqrt{4\pi}} \Pi_k \cg l 0 k 0 l 0 \right]}^2 \right] \nn 
&=& \frac{2}{\Pi^2_l} \left[ C_l(C_l - 2 l(l+1)C_l R) +2 C_l \sum_{k l_1} C^{\psi}_k C_{l_1} \left[ \frac{F(l_1 kl)}{\sqrt{4\pi}} \frac{\Pi_k\Pi_{l_1}}{\Pi_l} \cg l 0 k 0 {l_1} 0 \right]^2 + 4 C^2_l \sum_{k} C^{\psi}_k\left[ \frac{F(lkl)}{\sqrt{4\pi}} \Pi_k \cg l 0 k 0 l 0 \right]^2 \right] \nn 
&=& \frac{2}{2l+1} \left[ \tilde{C}^2_l + 4 C^2_l \sum_{k} C^{\psi}_k\left[ \frac{F(lkl)}{\sqrt{4\pi}} \Pi_k \cg l 0 k 0 l 0 \right]^2 \right]  \; .
\ea
\end{widetext}
\newpage
\def\urlprefix{}
  \def\url#1{}
\bibliography{lensbips}
\bibliographystyle{apsrev}
\end{document}